\documentclass[sigconf,anonymous=False]{acmart}

\usepackage{booktabs} %

\usepackage{lipsum}
\usepackage{hyperref}
\usepackage{tikz}
\usepackage{pgfplots}
\usepgfplotslibrary{colormaps}
\usetikzlibrary{pgfplots.colormaps}
\usetikzlibrary{pgfplots.dateplot}
\usetikzlibrary{patterns}
\usepackage{filecontents}
\usepackage{graphicx}
\usepackage{subcaption}
\usepackage{color, colortbl}
\usepackage{arydshln}
\usepackage{float}
\usepackage[utf8]{inputenc}
\usepackage{arabtex}
\usepackage{utf8}
\usepackage{units}
\usepackage{enumitem}
\usepackage{balance}
\usepackage{microtype} %
\usepackage[export]{adjustbox}
\usepackage{hyperref}
\setcode{utf8}

\definecolor{Gray}{gray}{0.9}

\newcommand{\afblock}[1]{\noindent{\textbf{#1 }}}
\newcommand{\takeaway}[1]{\noindent{\textbf{Findings.}} \textit{#1}}
\newcommand{\takeawayDeux}[1]{\noindent{\textbf{Takeaway.}} \textit{#1}}

\newcommand{\intents}{\mathcal{I}}
\newcommand{\topics}{\Theta}

\newcommand{\ie}{i.e.,\,}

\newcommand{\eg}{e.g.,\,}
\newcommand{\Eg}{E.g.,\,}
\newcommand{\etal}{et~al\@ifnextchar.{}{.\@}}
\newcommand{\etc}{etc\@ifnextchar.{}{.\@}}

\newcommand{\wrt}{w.r.t.\@}
\newcommand{\sref}[1]{\S~\ref{#1}}

\newcommand\encircle[1]{%
	\tikz[
		baseline={([yshift=-8pt]current bounding box.north)}
	]
		\node (X) [draw, shape=circle, inner sep=0, fill=black, text=white,scale=0.7] {\strut #1};
	\hspace{-4pt}
}

\AtBeginDocument{%
  \providecommand\BibTeX{{%
    \normalfont B\kern-0.5em{\scshape i\kern-0.25em b}\kern-0.8em\TeX}}}

\setcopyright{acmcopyright}
\copyrightyear{2022}
\acmYear{2022}
\acmDOI{10.1145/3487553.3524644}

\acmConference[WWW '22 Companion] {Companion Proceedings of the Web Conference 2022}{April 25--29, 2022}{Virtual Event, Lyon, France.}
\acmBooktitle{Companion Proceedings of the Web Conference 2022 (WWW '22 Companion), April 25--29, 2022, Virtual Event, Lyon, France}
\acmPrice{15.00}
\acmISBN{978-1-4503-9130-6/22/04}

\begin{document}

\title{Anonymous Hyperlocal Communities: What do they talk about?}
\author{Jens Helge Reelfs}
\email{reelfs@b-tu.de}
\affiliation{%
  \institution{Brandenburg University of Technology}
}
\author{Oliver Hohlfeld}
\email{hohlfeld@b-tu.de}
\affiliation{%
  \institution{Brandenburg University of Technology}
}

\author{Niklas Henckell}
\email{niklas@jodel.com}
\affiliation{%
  \institution{The Jodel Venture GmbH}
}

\renewcommand{\shortauthors}{Reelfs, et al.}

\begin{abstract}
In this paper, we study what users talk about in a plethora of independent hyperlocal and anonymous online communities in a single country: Saudi Arabia (KSA).
We base this perspective on performing a content classification of the Jodel network in the KSA.
To do so, we first contribute a content classification schema that assesses both the intent (why) and the topic (what) of posts.
We use the schema to label 15k randomly sampled posts and further classify the top 1k hashtags.
We observe a rich set of benign (yet at times controversial in conservative regimes) intents and topics that dominantly address information requests, entertainment, or dating/flirting.
By comparing two large cities (Riyadh and Jeddah), we further show that hyperlocality leads to shifts in topic popularity between local communities.
By evaluating votes (content appreciation) and replies (reactions), we show that the communities react differently to different topics; \eg{} entertaining posts are much appreciated through votes, receiving the least replies, while beliefs \& politics receive similarly few replies but are controversially voted. \end{abstract}

\maketitle

\section{Introduction}

Anonymity on internet platforms is often controversially discussed between \emph{i)} enabling freedom of speech and \emph{ii)} enabling toxic environments~\cite{papasavva2020raiders}.
Prior work studied the spectrum of discussed topics on anonymous to non-anonymous platforms showing that users have preferences which posts should be anonymous and which should not~\cite{correa2015many}.
The anonymous location-based app YikYak can be characterized with entertaining and informational contents leveraging self-supervised learning~\cite{black2016anonymous}, while others find evidence for flirting \& dating~\cite{Wu2017TPU} via crowdsourcing.
Qualitative studies identify a broad range of motivations for anonymous posts, \eg{} social isolation, social venting, requesting and granting emotional support, identity, while eliminating fear of rejection, to name a few~\cite{vatelausyikyak}.

This opens the question if anonymity yields to a richer content spectrum, especially in more conservative regimes.
In the case of Saudi Arabia, \cite{guta2015veiling} report on interviews with KSA women about boundaries and new freedoms, granted through the Internet---rendering anonymous platforms specifically interesting.
Alsanea protrays the Saudi life in the 2007's novel \emph{Girls of Riyadh}~\cite{girlsOfRiyadh} through the eyes of four young girls.
Nonetheless, society is continually changing, and has changed, \eg{} women's right to vote in 2015, or the KSA was about to lift the women driving ban in 2017.
Recent research empirically shows the Saudi Arabian user base to be active in creating and reacting to content over voting---contrasting a Western counterpart~\cite{reelfs2022pam}.
We take this as motivation to study social media posts in the KSA as intra-country and platform study.

Orthogonal to anonymity, recent online messaging platforms embrace hyperlocality, \ie{} they display posted content only to spatially local users.
It is an open question whether this property implies shifts in discussed contents.
One platform that combines both properties---anonymity and hyperlocality---is Jodel.
The app only displays content posted within the users' proximity---unlike Twitter and other platforms, no communication with remote users is possible.
The platform became popular in the KSA in 2017~\cite{reelfs2022pam}.

\afblock{Research Questions.}
Given the different cultural background in the KSA, we are interested in (RQ1) what are discussed contents on the Jodel platform in the KSA and how is it perceived \& reacted upon by the communities.
We thereby study effects of Jodel's key design features of \emph{anony\-mity} and \emph{hyper\-locality}.
What are the Jodel KSA users talking about?---\emph{Behind the veil}.
Subsequently, we raise the question (RQ2): 
How can we design a suitable crowdsourcing annotation schema to assess Jodel content?
Last, (RQ3) how can we classify hashtags---as proxy measure for post content.

\afblock{Methodology.}
We take the rare chance to analyze ground truth information provided by the social network operator to study a random sample of Jodel content posted within the KSA.
We enable content classification using a new content annotation schema that assesses the \emph{intent} (why) and the \emph{topics} (what) of a post.
We apply the scheme to 15k randomly sampled posts that are annotated by expert native-speaking classifiers.
By splitting the data set by city, we study local content biases between two major cities in the KSA.
Leveraging empirical data, \eg{} vote scores, or \#replies, we study how users appreciate discussed content-classes.

In a last step, we extract and classify hashtags \wrt{} sensitivity.

\noindent{
Our \textbf{contributions} are as follows.
}
\begin{itemize}[leftmargin=*]
    \item We contribute a content classification schema to classify social media posts by their intent (why) and topic (what).
    \item We apply the schema to ground truth data from the hyperlocal and anonymous Jodel network in the KSA, using native-speaking expert classifiers.
    We successfully put this schema to work for classifying Jodel content (\eg{} high rater agreement).
    In contrast to what might be expected in anonymity, we observe a rich set of benign (yet at times controversial in conservative regimes) but non-toxic topics.
    We further show local topical biases within the two largest cities.
    Differences exist also in how content triggers responses \& how users appreciate content through voting. 
    \item We develop a second schema to classify the top 1k hashtags.

\end{itemize}

\begin{figure}[t]
	\centering
	\includegraphics[width=0.99\linewidth]{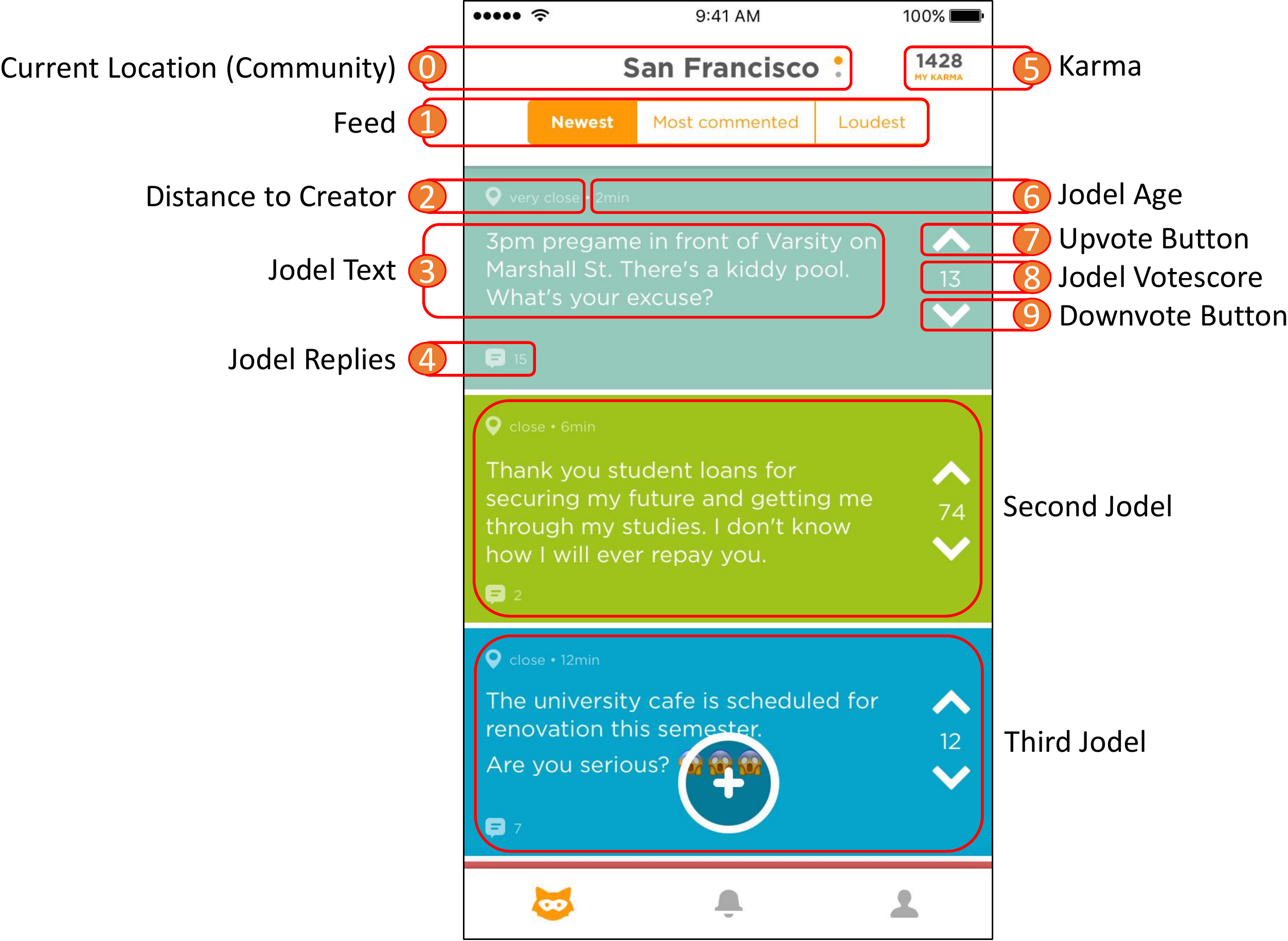}
	\caption{
		Jodel iOS mobile application.
		\vspace*{-1em}
	}
	\label{fig:jodelapp}
\end{figure}

\section{Jodel - Hyperlocal Messaging App}
	\label{sec:jodelapp}
	Jodel
	is a mobile-only messaging application which we show in Figure~\ref{fig:jodelapp}.
	It is \emph{location-based} and establishes local communities relative to the users' location \protect\encircle{0}.
	Within these communities, users can {\em anonymously} post both images and textual content of up to 250 characters length \protect\encircle{3} (\ie{} microblogging) and reply to posts forming discussion threads \protect\encircle{4}.
	Posted content is referred to as ``Jodels'', colored randomly \protect\encircle{3}.
	Posts are only displayed to other users within close (up to $\approx$20km) geographic proximity \protect\encircle{2}.
	Unlike other networks (\eg{} Facebook, Twitter, etc.), communication to remote users outside the local community is \emph{not} possible.
	This biases all posts to having relevance to the local community---a very different mode of communication that opens questions on \emph{what} is being posted and if content differs between communities.

	Further, all communication is {\em anonymous} to other users since the app does not display any handle or user-related information.
	Only {\em within} a single discussion thread, Jodel enumerates users to enable referencing other users.
	The app builds upon three main feeds showing \protect\encircle{1}: i) most \emph{recent} threads, ii) most \emph{discussed} threads, and iii) \emph{loudest} showing threads with highest voting scores (likes).

	Being a key success parameter for an anonymous messaging app, Jodel employs a community-driven filtering and moderation scheme to avoid harmful or abusive content.
	In Jodel, content filtering relies on a distributed voting scheme in which every user can increase or decrease a post's vote score by up- (+1) \protect\encircle{7} or downvoting (-1) \protect\encircle{9} (similar to StackOverflow).
	Posts reaching a cumulative vote score \protect\encircle{8}\,below a negative threshold \mbox{(\eg{} -5)} are no longer displayed.
	Depending on the number of vote-contributions, this scheme filters out adverse content while also potentially preferring mainstream content.
	To increase user engagement \wrt{} posting and voting, Jodel uses lightweight gamification by awarding \emph{Karma} points \protect\encircle{5}.

\subsection{Data Set Description and Ethics}
	\label{sec:Dataset_Description_and_Statistics}

	\afblock{Meta Data.}
	The Jodel operator provided us with excerpts of {\em anony\-mized} content data of their network.
	The obtained data contains \emph{metadata} for 469\,M posts created within the KSA by 1.3\,M users.
	It spans over multiple years from the beginning of the network in 2014 up to August 2017.
	Due to being captured for operational purposes during the rollout, more recent data is not available to us.
	To protect user privacy on raw-data level, it is limited to metadata only without textual content, and anonymized user IDs.
	The structure of this dataset includes 3 categories: interactions, content, and users.

	Defining ``a'' community is not possible on Jodel given that the app always displays content relative to each user's location and thus usually differs from user to user. 
	That is, every user might experience a slightly different community to interact with, which cannot be reconstructed from the data. 
	To solve this, we assign each interaction to a nearby major city or district, which generates clusters of interactions that we refer to as communities.

	\afblock{Smaller sub-dataset with textual posts.}
	For \emph{content} classification, we have been provided with a random sample of 15\unit{k} threads throughout the country.
	Furthermore, we have access to used hashtags counts from a \#1.12\unit{M} random subsample.
	Users generally post content consciously into the public application domain.
	Albeit against the Jodel Terms of Service that forbid posting personal information, \ie{} identifying individuals, posts may still contain such information or even references to other social media platforms directly opening up the possibility for identification of individuals.
	In liaison with Jodel and our university's Chief Data Officer, our crowdsourcing experiments \emph{cannot} be conducted by any personnel not associated to our university; \ie{} to protect users' privacy, we cannot use typical crowdsourcing platforms, such as Amazon Mechanical Turk or Microworkers, but must employ coders under direct governance.
	We store data strictly encrypted on a dedicated firewalled server; (selected) content is only accessible to authenticated internal users (coders).
	We generally inform and synchronize with the Jodel operator on analysis we perform on their data.

\section{Content Classification Schema}
\label{sec:design}

\begin{table}

  \centering
  \smaller
  \resizebox{\linewidth}{!}{
    \begin{tabular}{llrr}
        Abbreviation & Class Information & \#Annot    & pdf \\ \toprule
        \multicolumn{1}{l}{$\intents{}=$ \textbf{Intents}} & \multicolumn{3}{r}{\#Coders=2; $\alpha_{M}^\intents{}=0.74$, \emph{substantial}} \\ \midrule
      EntObserv   & Entertaining Observation                    
          &   122 & $0.019$ \\
      DistRelComp & Distress Release \& Complain. ab. Self      
          &   580 & $0.091$ \\
      GenEntert   & General Entertainment                       
          &   618 & $0.097$ \\
      Info   & Information Sharing                         
          &   638 & $0.100$ \\
      SocVentComp & Social Venting \& Complain. ab. Others      
          &   644 & $0.101$ \\
      Other       & Other                                       
          &   691 & $0.108$ \\
      Self   & Self Expression                             
          &   927 & $0.146$ \\
      Seek   & Seeking Interaction/Information             
          & 2,149 & $0.337$ \\
      \midrule
      $|\intents{}|=8$ & \multicolumn{1}{r}{$\Sigma$} & 6,191 & 1.0\\
      \toprule
      \multicolumn{1}{l}{$\topics{}=$ \textbf{Topics}} & \multicolumn{3}{r}{\#Coders=2; $\alpha_{M}^\topics{}=0.64$, \emph{substantial}} \\ \midrule
      IllegalViolence & Illegal \& Violence         
          &   155 & $0.018$ \\
      AnimalsNature   & Animals \& Nature           
          &   228 & $0.026$ \\
      FitnessHealth   & Fitness \& Health           
          &   267 & $0.031$ \\
      FoodDrink       & Food \& Drink               
          &   358 & $0.042$ \\
      ProdService     & Products \& Services        
          &   376 & $0.044$ \\
      EduWork         & Education \& Work           
          &   479 & $0.056$ \\
      FashionBeauty   & Fashion \& Beauty           
          &   506 & $0.059$ \\
      SocialMedia     & Social Media                
          &   608 & $0.071$ \\
      BeliefsPol      & Beliefs \& Politics         
          &   707 & $0.082$ \\
      EntertCulture   & Entertainment \& Culture    
          &   896 & $0.104$ \\
      Other               & Other                       
          &   983 & $0.114$ \\
      Self            & Self (Personal)             
          & 1,092 & $0.127$ \\
      PeopleRelation  & People \& Relationships     
          & 1,960 & $0.228$ \\
      \midrule
      $|\topics{}|=13$ & \multicolumn{1}{r}{$\Sigma$} & 8,615 & 1.0\\
      \bottomrule
  \end{tabular}  
  }
  \caption{
    Annotation Schema \& Crowdsourcing Results. 
    We code \emph{intents} ($\intents$) catching the individual incentives, and \emph{topics} ($\topics$) representing discussed contents. 
    Figures are given in the amount of jodels, number of annotations, and the resulting probability. 
    The overall coder agreement by MASI distance is substantial, $\alpha_{M}^\intents{}=0.64$ and $\alpha_{M}^\topics{}=0.74$.
    \vspace*{-3em}
  }
  \label{tab:schema}
\end{table}

In this section, we contribute a crowdsourcing schema that enables to classify content posted in social media platforms.
Our schema assesses two key aspects: \emph{i)} why a user posted, \ie{} what is the purpose or intent(s) $\intents{}$ of a post.
\emph{ii)} What topics $\topics{}$ are presented in a post.
For each the intents $\intents{}$ and the topics $\topics{}$, multiple labels can be attributed by human annotators to a single post.

\afblock{Design objectives and development.}
We iteratively developed and refined the presented schema over multiple months.
Our objective was to arrive at a \emph{minimal set of categories} easing the classification task.
The categories should have little to no semantic overlap to make classes easily distinguishable; for both easing annotation and better interpretability of results.
Categories must further be sufficiently expressive for the content posted on Jodel, \ie{} the amount of posts annotated with ``Other'' should be minimal.
The design of the categories naturally involves a trade-off between being very specific (many categories) and ease of use (few categories).

\afblock{Intent $\intents{}$ of a post.}
In the first category, we assess \emph{why} users post in social media, \ie{} the user's driving intent of a post as interpreted by the human annotator.
In our schema, we use eight possible intents that we base upon a prior work's~\cite{strangers_on_your_phone} taxonomy, derived from semi-structured interviews with social media users.
Table~\ref{tab:schema} shows the list of intents $\intents{}$, 
\eg{} if a user is sharing information or is seeking for information.
The selected intents can be assessed by human classifiers solely by reading the posted textual content. %
As posts may have multiple intents, we allow multi-labels per post.

\afblock{Topic $\topics{}$ of a post.}
In the second category, we assess \emph{what} topic a post is about.
Our initial set of categories bases on prior work on content classification of the Whisper network~\cite{paul2011question, correa2015many}, which we iteratively refine and adapt to content shared via Jodel KSA.
We show the list of topics $\topics{}$ in Table~\ref{tab:schema}.
We also opted for multi-labelling.

\afblock{Iterative schema development.}
We based the initial version of the schema on prior work~(\cite{strangers_on_your_phone} for the intents and ~\cite{paul2011question, correa2015many, Wu2017TPU} for the topics), that we have iteratively refined and adapted in multiple classification campaigns, each based on a small random samples of Jodel posts.
Qualitative coder feedback was in line mentioned works, we do not find any specifically toxic environment.
An empirical view shows that from the overall KSA's content we find only p$\approx$1.6\% of outvoted (disliked) or otherwise blocked posts and replies.
Though this figure is higher for in-app prominently displayed posts with p$\approx$7.3\%, only a minor fraction of these posts has been blocked by escalated moderation and flagging with p$\approx$2.3\%.
This indicates that applied moderation works; especially due to distributed moderation being very interesting by itself, we leave this topic for future work.
That is, we excluded any focused toxicity class as it also arguably might not fit well into topics nor intents.

In each campaign, we identified categories being used seldom, missing, or being semantically ambivalent, \ie{} they resulted in strong disagreement among annotators.
After each classification run, we discussed disagreement and other challenges with our annotators, ultimately leading to an improved version of the schema.
We present our final schema and put it to work enabling us to classify Jodel posts with substantial coder agreement.

\begin{figure}[t]
  \centering
  \includegraphics[width=.78\linewidth]{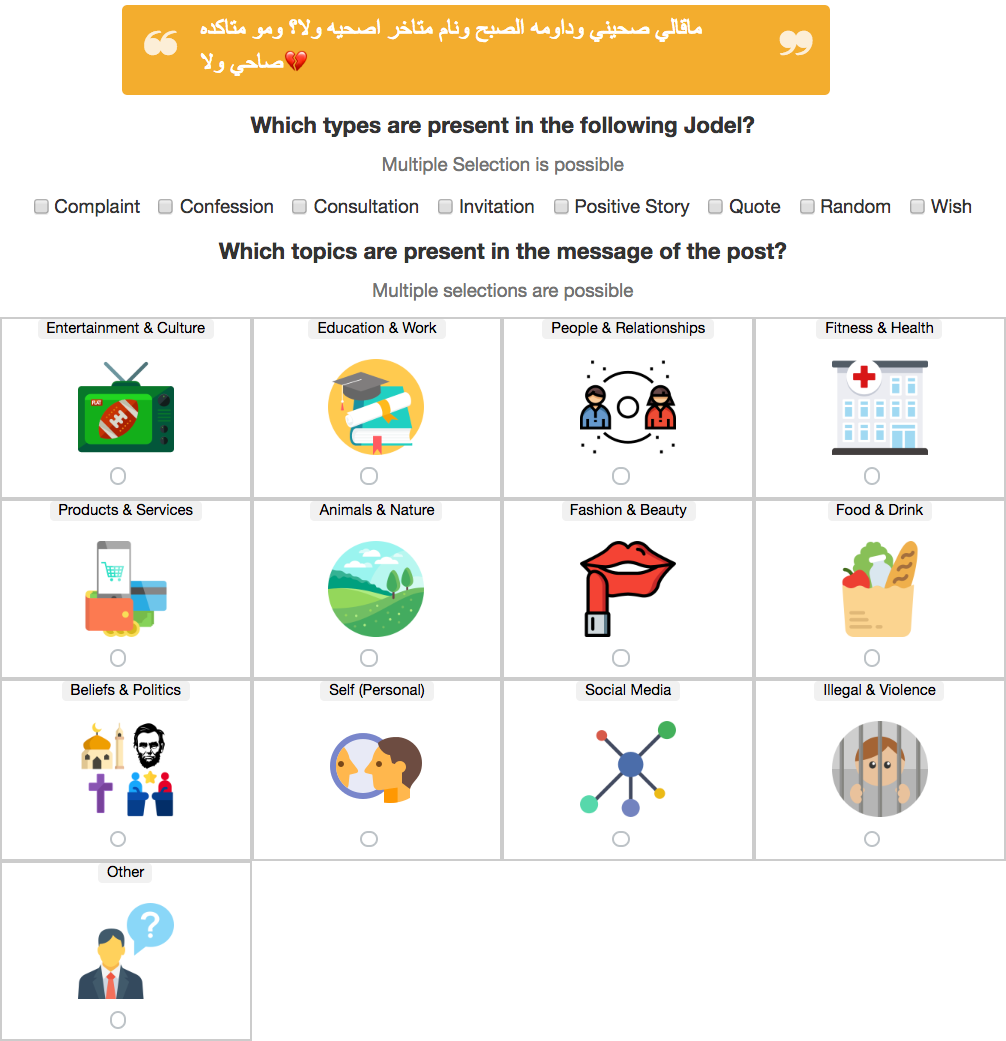}
  \vspace{-1em}
  \caption{
      Crowdsourcing Classification System. Our coders are presented a post to be read. Then, they answer two subsequent questions: \emph{i)} What is the intent, and \emph{ii)} What is the topic of this post?---Allowing multiple labels.
      \vspace{-1em}
  } 
  \label{fig:sys}
\end{figure}

\afblock{Implementation.}
We realized the crowdsourcing system from scratch as a web application in PHP, shown in Figure~\ref{fig:sys}.
The system is self-hosted and enables us classifying sensitive content that cannot be made available to external services or users (\eg{} via common crowdsourcing platforms such as Amazon Mechanical Turk).
We can define and monitor annotation campaigns: Describe which posts should be classified by how many of the available annotators.

\section{Classification Campaign}
\label{sec:crowdsourcing}

\subsection{Study Design}
\label{sec:studydesign}

We apply our content classification schema (\sref{sec:design}) to annotate Jodel posts in the KSA (see \sref{sec:Dataset_Description_and_Statistics} for the data set description).
To protect the users' privacy and in compliance with the Jodel ToS, we cannot share the posts on external platforms such as Amazon Mechanical Turk.
We thus run all campaigns on internal and protected machines that are only accessible to expert classifiers that we invite and associate with our research group.
The human annotators are experts that \emph{i)}~are Arabic native speakers and \emph{ii)} are familiar with the dialect spoken in the KSA (\eg{} by originating or having lived in the region).
Within our schema development, we realized that the annotators' origin (\eg{} Egypt) \emph{can} challenge the understanding of local KSA dialects, and can thus lead to disagreement between annotators.
Therefore, we selected future annotators by removing language boundaries \& ensuring more consistent annotations.
Using expert classifiers reduces the number of classifiers needed; prior work showed that using non-expert classifiers requires a factor of 4 more classifiers~\cite{snow2008cheap}.
Also, since all of our classifiers are known to us and trusted, we do not need to employ control questions to detect cheating attempts as in crowdsourcing on public platforms.

For coherent classification results, we focus on the content of the starting post, \ie{} not classifying complete threads nor replies, due to findings within the schema development phase.
We experimented with presenting more contextual information to our annotators by including the entire discussion thread (\ie{} original post \emph{and} its replies).
Particularly longer discussion threads tend to shift from one topic to others and are thus challenging to label coherently.
For the schema development, we employed four expert classifiers aged 20-30 years with a 1:1 male:female ratio.
Three out of four classifiers have prior experience with the KSA dialect (\eg{} from having lived in Saudi Arabia or Oman).
Within development and schema optimization, we performed about 7,700 classifications across various setups with multiple classifiers into a feedback loop of discussing ambiguity, disagreement, ambivalence, and other experiences---each resulting in a new schema version.
Later we settled with the final schema that includes topics and intents.

\subsection{Campaign \& Schema Quality Evaluation}

To study the content of Jodel posts in SA with our fixed final schema,
we employed two classifiers (aged 20-30 years, male and female, from Syria and Iraq), who iteratively performed five subsequent classification campaigns (Table~\ref{tab:final_campgains}).
All campaigns use sampled post data from \emph{i)} the entire KSA, \emph{ii)} Jeddah, and \emph{iii)} the capital Riyadh.
They first completed a training period to qualitatively familiarize with Jodel contents and our schema.
Since the agreement for all campaigns is high, we opted for using all campaigns for evaluation.

Next, we evaluate the quality of the campaign and the schema.

\afblock{\emph{Qualitative Coder Feedback.}}
From analyzed Jodels, both annotators believe that dominantly teenagers and young people use the platform.
Since the Jodel network is anonymous, we lack any demographic info to validate this claim.
The classifiers further noted that a number of posts focus on finding partners for online games, especially \emph{ludo star}, a mobile app version of the board game \emph{Don't Get Angry}~\cite{MenschAergereDichNicht} (identified as \emph{seeking interaction} in the later results).

\afblock{Coder Agreement.}
We measure our expert classifier interrater-agreement with Krippendorf's alpha~\cite{krippendorff2018content}. %
A standard approach that provides several benefits: \emph{i)} it behaves well with any number of classifiers, \emph{ii)} it is capable of handling missing data, \ie{} single classifications, \emph{iii)} it adjusts for sample sizes, and \emph{iv)} it may be used for various types of data--nominal in our case.
Due to our \emph{multi-\-label} approach, we further need to use a suitable distance metric that compares sets of labels.
We present our agreement results in Table~\ref{tab:final_campgains} using various distance metrics: Binary, Jaccard, and MASI~\cite{passonneau-2004-computing}.

There is no clear up-/downwards trend in agreement across the campaigns; thus, it remains unclear whether annotators accustom better to the classification scheme.
When analyzing intents and topics separately, we note that intents $\intents{}$ generally suffer less from non-agreement.
The topic $\topics{}$ classification has led to less agreement consistently.
Later iterations yield better results for both, intents and topics.
Our results should be viewed with care as \cite{krippendorff2018content} suggests not to use data sets with alpha values below $0.667$ for any non-tentative conclusion, yet our achieved agreement is still well above chance.

\begin{table}

    \small
    \centering
    \resizebox{\linewidth}{!}{
    \begin{tabular}{rrrcrrrl}
        \toprule
        {No}        & {\#Posts} & {\#Coders>1} & {$\intents{}/\topics{}$} & {$\alpha_B$} & {$\alpha_J$} & {$\alpha_M$} & Agreement\textsubscript{$M$}\\
        \midrule
        1                                                     & 733     & 0     &              & -      & -      & -      & \\
        2                                                     & 1,999   & 0     &              & -      & -      & -      & \\
        \midrule
        3                                                     & 400     & 398   & $\intents{}$ & $0.66$ & $0.66$ & $0.66$ & substantial\\
                                                              &         &       & $\topics{}$  & $0.44$ & $0.57$ & $0.52$ & moderate\\
        4                                                     & 1,000   & 993   & $\intents{}$ & $0.69$ & $0.69$ & $0.69$ & substantial\\
                                                              &         &       & $\topics{}$  & $0.45$ & $0.60$ & $0.55$ & moderate\\
        5                                                     & 400     & 398   & $\intents{}$ & $0.70$ & $0.70$ & $0.70$ & substantial\\
                                                              &         &       & $\topics{}$  & $0.57$ & $0.62$ & $0.61$ & substantial\\
        \midrule

        {all}                                                   & 4,532   & 1,789 & $\intents{}$ & $0.74$ & $0.68$ & $0.74$ & substantial\\
                                                              &         &       & $\topics{}$  & $0.57$ & $0.68$ & $0.64$ & substantial\\
        \bottomrule
    \end{tabular}
    }
    \caption{
        Classification agreement for multiple iterations
        on thread starting posts;
        For intents $\intents{}$ \& topics $\topics{}$
        we show classifier agreement by $\alpha_B$ Binary, $\alpha_J$ Jaccard, and $\alpha_M$ MASI distance. 
        There is a substantial agreement between our two coders. 
        Coders agree better on intents ($\mathcal{I}$) than topics ($\topics$). 
        \vspace*{-3em}
    }
    \label{tab:final_campgains}
\end{table}

\afblock{Multi-Labels.}
We analyzed the amount of classified posts with multiple intents $\intents{}$ or topics $\topics{}$.
While $\intents$ intents almost accidentally only were assigned a multi-label twice,
we found 1,917 multi-labels across our coders for $\topics$ topics ($p$=\{2: 0.81, 3: 0.17, 4: 0.02\}).
Observed multi-labels are usually not specifically in line between the classifiers as can be seen in lower $\alpha_B$ scores across the board; lower $\alpha_M$ values in comparison to $\alpha_J$ confirms this finding as the MASI distance adds a distinct bias according to subset-similarities.

Thus, we also investigated on observed confusion between multi-label $\topics$ topic annotations and annotators.
We find a strong diagonal as expected due to substantial agreement.  %
However, we identify the axis along \emph{People \& Relationships} as most ambiguous. %
Other single $\topics$-hotspots are worth a look: Some may raise self-explainable confusion: \eg{} $\topics$-\emph{Self} $\times$ \(\emph{FitnessHealth}, \emph{EduWork}, \emph{FashionBeauty}\), or \emph{EntertCulture} $\times$ \emph{AnimalsNature}.

\afblock{Overall Confusion.}
In Figure~\ref{fig:confusion_intents} and Figure~\ref{fig:confusion_topics}, we provide the complete picture of confusion within our annotation schema for intents and topics.
We de-biased the join operation by introducing a natural weighting factor of $n \times m^{-1}$ as it would otherwise favor multi-label annotations.
Note the log color scale.

For \emph{intents} $\intents{}$, in Figure~\ref{fig:confusion_intents}, we observe a strong correlation across the diagonal as expected from substantial annotator agreement.
However, several confusion hotspots remain, some being self-ex\-pla\-na\-tory: \Eg{} \emph{GenEntert} $\times$ \emph{EntObserv}, or \emph{DistRelComp} $\times$ \emph{SocVentComp}; yet we must take note of the rest.

As for \emph{topics} $\topics{}$, in Figure~\ref{fig:confusion_topics}, we observe similar patterns to the multi-label confusion.
Though we observe substantial agreement on the strong diagonal, we again see evidence for ambiguity along \emph{PeoplRelation}, against \emph{FitnessHealth}, and naturally along $\topics$-\emph{Other}.
Noteworthy, we also identify several hotspots along $\topics$-\emph{Self}.
\takeawayDeux{ %
    We have presented our schema of intent and topics at work within a series of subsequent crowdsourcing campaigns for Jodel KSA.
    Within about 15\unit{k} annotations (intent-\#annot=6,191, topic-\#annot=8,615), we find substantial coder agreement ($\alpha_M^\intents{}=0.75$, $\alpha_M^\topics{}=0.64$) across \#posts=1,789 (\#Coders=2) and conclude that our proposed schema has sufficient quality for further evaluation.
    }

\section{What Jodel Users talk about in SA}
\label{sec:eval}

In this section, we study the Jodel post contents (\ie{} \emph{topics} and their \emph{intents}) in the KSA that result from our classification campaign.

\subsection{Countrywide Perspective on Jodel Content}

\begin{figure}[t]
    \centering
    \includegraphics[width=\columnwidth]{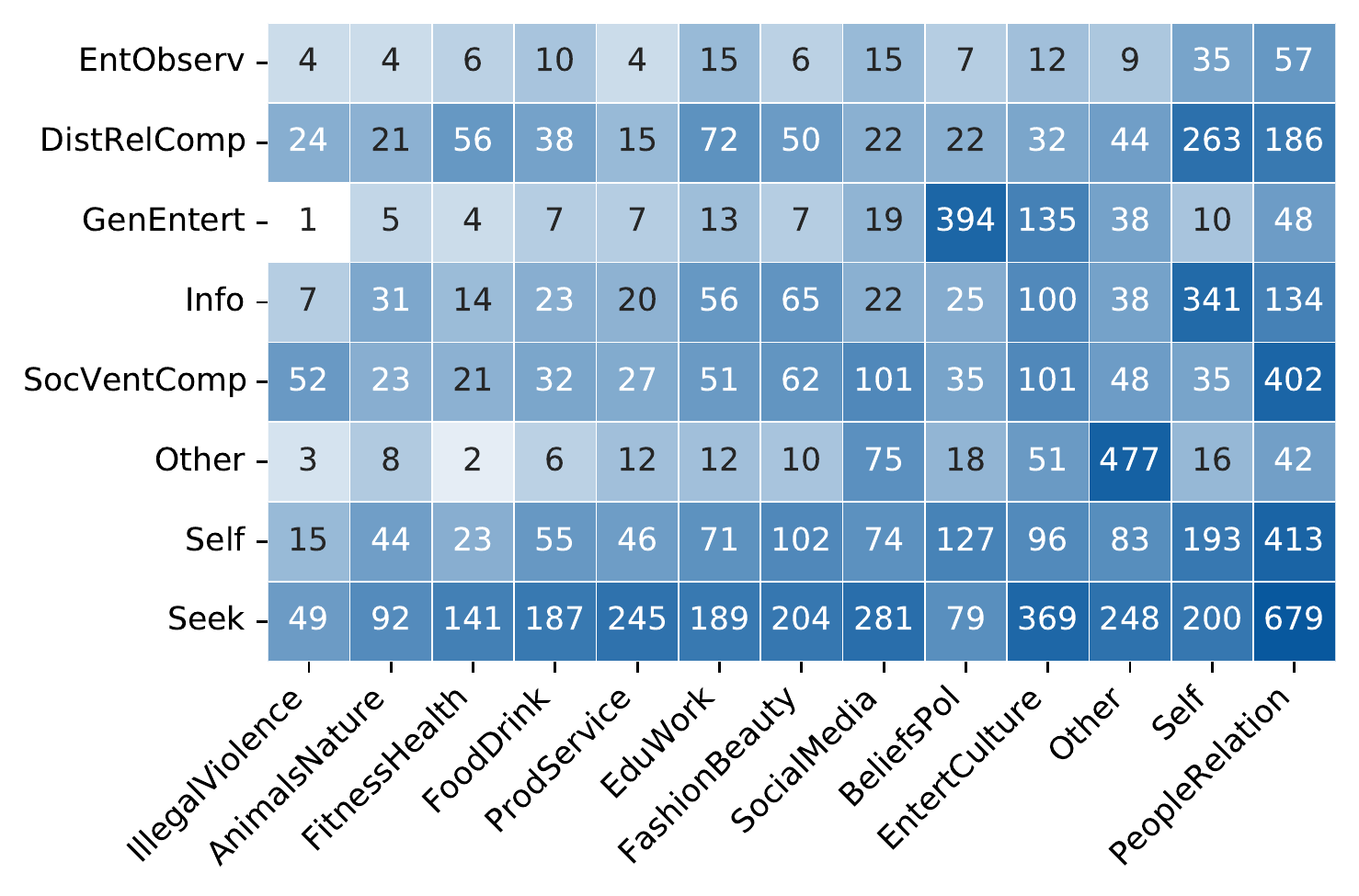}
    \vspace{-3em}
    \caption{
        Overall annotation counts, intents $\intents{} \times \topics{}$ topics.
		\vspace{-1.5em}
    }
    \label{fig:overall_counts}
\end{figure}

We begin with analyzing the content classification for the country-wide overall annotations, before we study differences between two cities.
First, we discuss Table~\ref{tab:schema} showing the popularity of topics $\topics{}$ and intents $\intents{}$ by annotation counts.
Since topics and intents are intertwined, we also show the combination of $\intents{} \times \topics{}$ as a heatmap in Figure~\ref{fig:overall_counts}.
We complement this heatmap by discussing topic distributions across intents next (not shown). %

\afblock{Intents $\intents{}$.}
The dominant intents are (see Table~\ref{tab:schema}): $\intents$-\emph{Seek} ($\Sigma$34\%) and \emph{Self} ($\Sigma$15\%), followed by \emph{Soc\-Vent} \& \emph{Dist\-Rel} ($\Sigma$19\%), \emph{Info} ($\Sigma$10\%), and \emph{Entertainment} ($\Sigma$12\%).
We only observe little disagreement between classifiers, explained by a possible ambiguity within $\intents$-\emph{Entobserv} $\times$ \emph{Gen\-Entert}, or apparent confusions along \emph{Self}.

\afblock{Topics $\topics{}$.}
Albeit slightly weaker annotator agreement, the discussed topics largely revolve around $\topics$-\emph{PeopleRelation} accounting for $\Sigma$25\% annotations, which is also our most confused category.
We find other popular themes in $\topics$-\emph{Self} ($\Sigma$13\%), \emph{Other} ($\Sigma$13\%), followed by \emph{EntertCulture} ($\Sigma$10\%), and \emph{BeliefsPol} ($\Sigma$8\%).
\emph{IllegalViolence} ($\Sigma$2\%), \emph{AnimnalsNature} ($\Sigma$3\%), and \emph{FitnessHealth} ($\Sigma$3\%) are least popular.

\afblock{Intents $\intents{} \times \topics{}$ Topics.} %
We identify specific hotspots of interests by combining $\intents{} \times \topics{}$ as a heatmap in Figure~\ref{fig:overall_counts}.
Jodel is mostly being used out of the intent of $\intents$-\emph{Seek}ing Information \& Interaction ($\Sigma$34\%) for $\topics$-\emph{People\-Relation} ($p[\topics|\intents]$=23\%) and \emph{Entert\-Culture} ($p[\topics|\intents]$=12\%), closely followed by others.
Likewise, finding $\intents$-\emph{Self} ($\Sigma$15\%) Expression across the board, users again focus on $\topics$-\emph{PeopleRelation} ($p[\topics|\intents]$=23\%) and \emph{Self} Expression ($p[\topics|\intents]$=14\%).
Out of $\intents$-\emph{Gen\-Entert} ($\Sigma$10\%), we want to highlight $\topics$-\emph{BeliefPol} ($p[\topics|\intents]$=57\%).
Where\-as $\intents$-\emph{Soc\-Vent\-Comp} ($\Sigma$10\%) almost naturally goes along with the topic $\topics$-\emph{People\-Relation} ($p[\topics|\intents]$=41\%),
$\intents$-\emph{Dist\-Rel\-Comp} ($\Sigma$9\%) aligns with $\topics$-\emph{Self} ($p[\topics|\intents]$=31\%) and \emph{People\-Relation} ($p[\topics|\intents]$=22\%).

\afblock{Anonymity.}
From our analysis, it comes apparent that most content posted on Jodel indeed is related to users' intent for $\intents$-\emph{Seek}ing Information \& Interaction, and \emph{Self} Exression accounting for 50\% of all annotations with strong trends towards the topics $\topics$-\emph{PeopleRelation}, \emph{Entert\-Culture}, and \emph{Self} totaling for $p[\topics|\intents]$=(37\%, 16\%, 13\%, $\Sigma$66\%) within these intents.
Further, another $\Sigma$19\% of posts are driven by $\intents$-\emph{SocVentComp} and \emph{DistRelComp} within the same topical regime.
Unfortunately, crowdsourcing a well-suited anonymity-sensitivity score relying on many (n=89) coders~\cite{wang2014whispers} is not possible in our case.
Foreshadowing our categorization on hashtags in~(\sref{sec:hashtags}), we nonetheless conclude that found prominent interaction situations are in line with the key design feature of being anonymous.
Individuals may find safety behind the veil of anonymity, allowing for free speech about personal experiences, wishes, questions, or possibly controversial opinions.

\afblock{Hyperlocality.}
Our generic intent $\intents{} \times \topics$ topic schema does not allow for a distinguished evaluation whether a post refers to \emph{anything} local, which turns out to be a challenging question.
With qualitative feedback from our annotators, we conclude that \eg{} a larger part of $\intents$-\emph{Seek}ing Interaction \& Information actually refers to local matchmaking, events, local services, or educational institutions.
By focusing on the platforms content and driving factors first, we leave distinguished analyses of a \emph{well-suited sense of locality} to future work.
Though, we discuss this topic in more detail within the next \sref{sec:hyperlocality} comparing intents \& topics in Jeddah $\times$ Riyadh, and provide deeper insights via hashtags in \sref{sec:hashtags}.

\takeaway{%
    ($\intents$) Seeking Information \& Interaction and Self Expression are the predominant drivers for creating content within the Jodel KSA communities.
    ($\topics$) Discussions and statements largely revolve around People \& Relationships and personal statements (Self).
    Besides a significant amount of entertaining content, Jodel's anonymity promotes personal and respectively potentially sensitive content (non-illegal).
}

\begin{figure*}[t]
    \centering
    \begin{subfigure}[t]{.3\linewidth}
        \centering
        \includegraphics[width=\columnwidth]{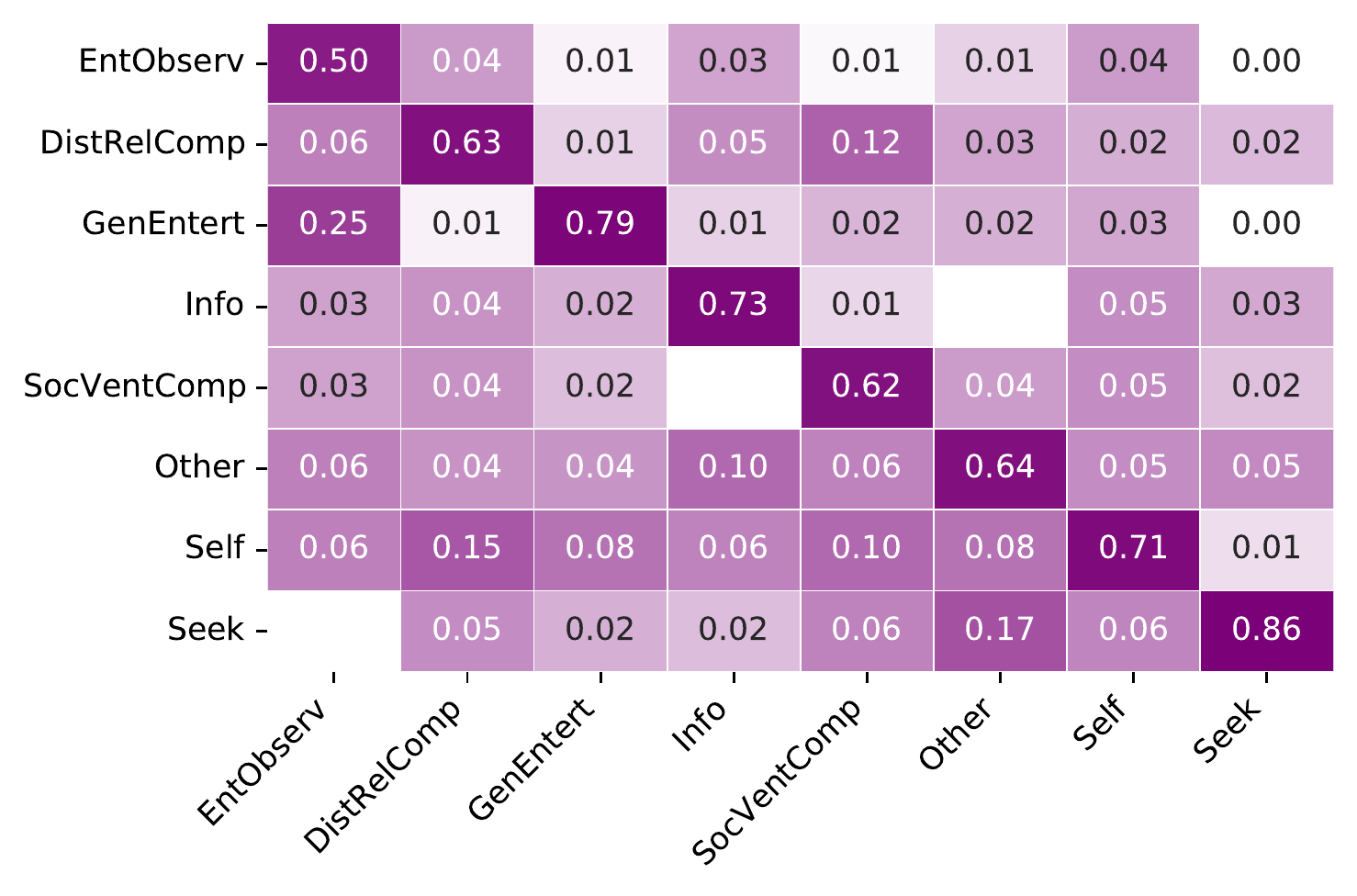}
        \caption{Relative Confusion between intents $\intents{}$.}
        \label{fig:confusion_intents}
    \end{subfigure}
    \begin{subfigure}[t]{.3\linewidth}
        \centering
        \includegraphics[width=\columnwidth]{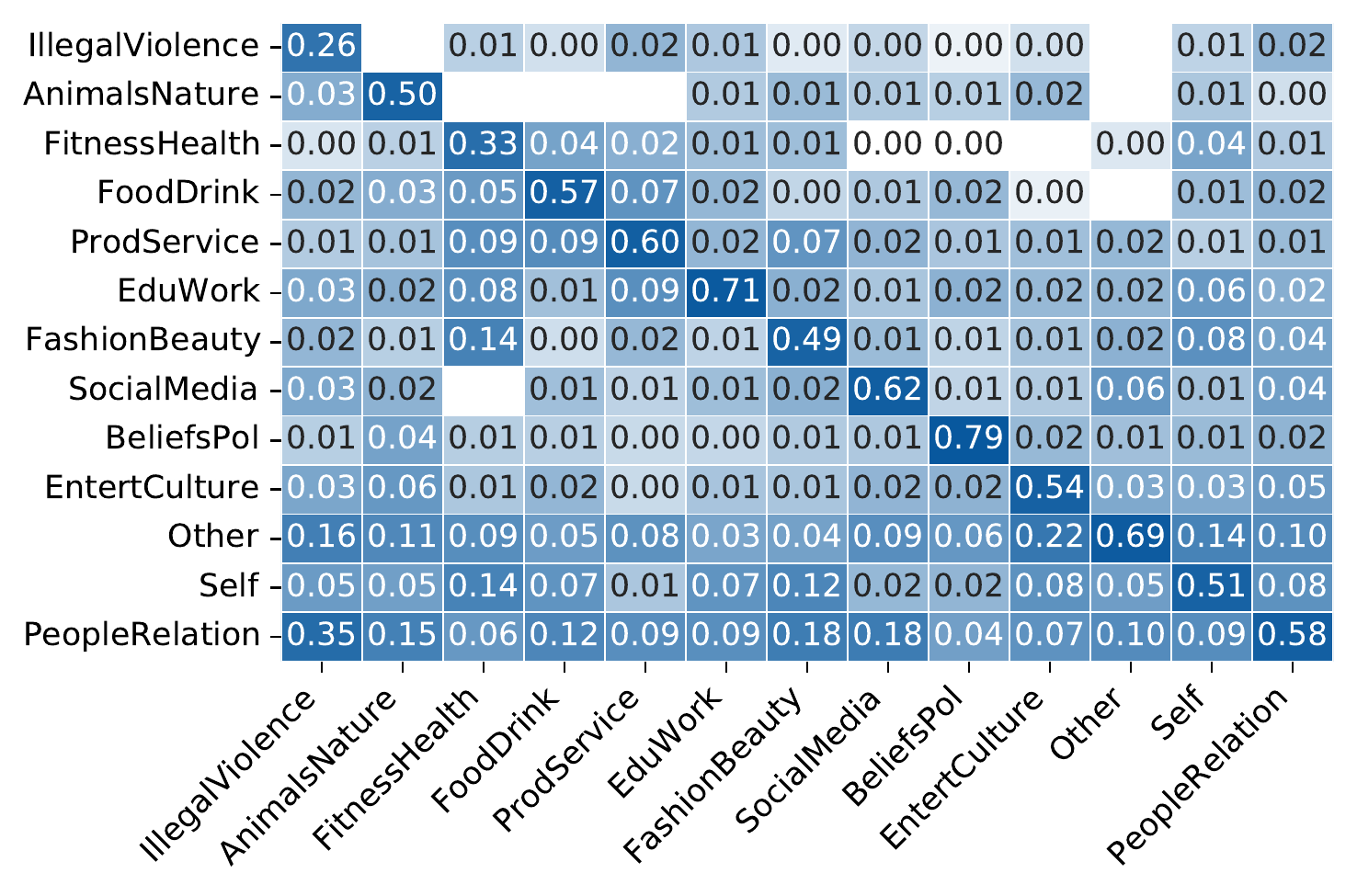}
        \caption{Relative Confusion between topics $\topics{}$.}
        \label{fig:confusion_topics}
    \end{subfigure}
    \begin{subfigure}[t]{.3575\linewidth}
        \centering
        \includegraphics[width=\columnwidth]{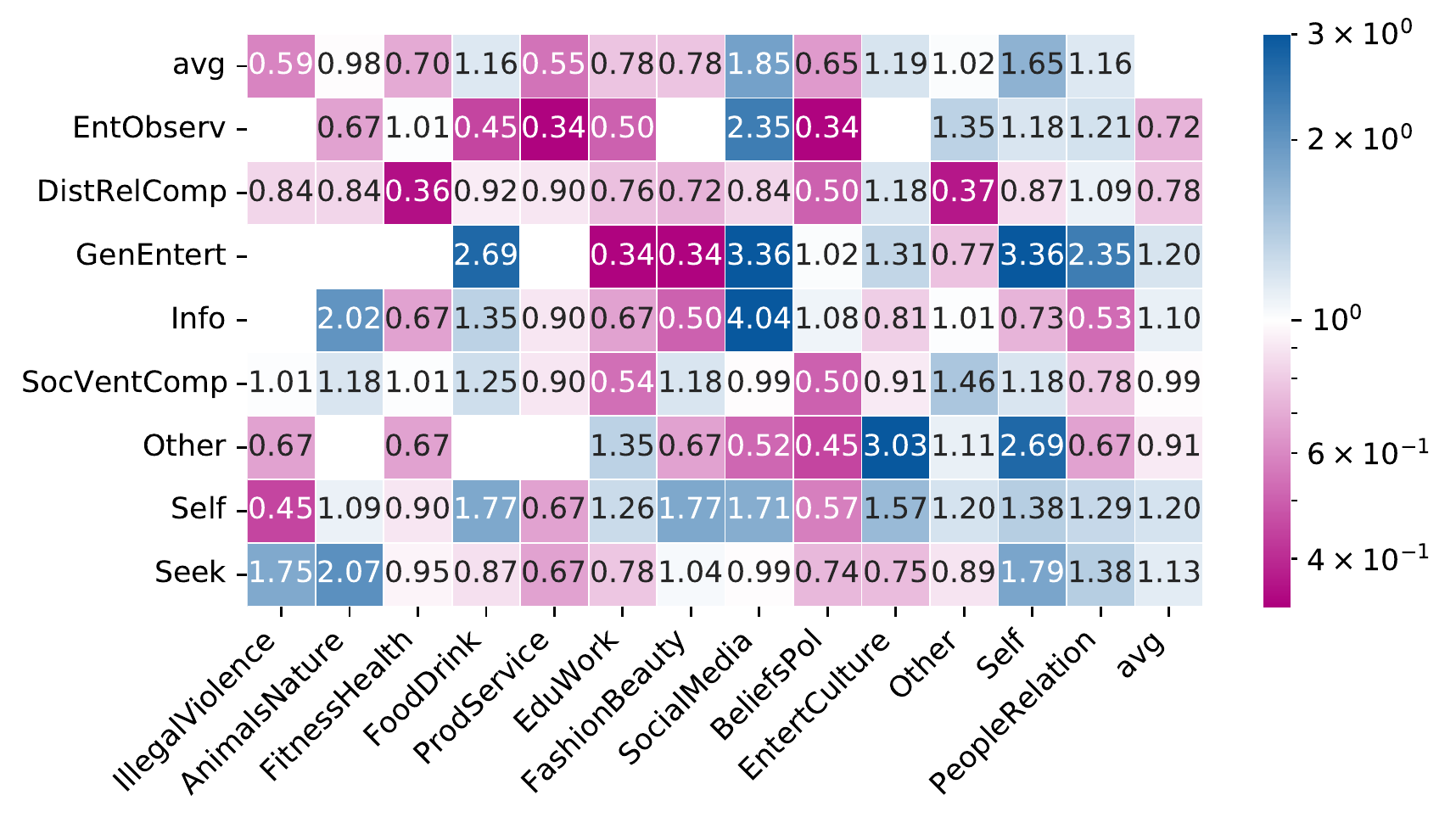} %
        \caption{
            Jeddah$\times$Riyadh PDF relative delta.
        }
        \label{fig:diffs_city}
    \end{subfigure}
    \caption{
        (Left, middle:) Relative Confusion, column wise normalization, sums to 100\%. 
        De-Biased multiset join by weighting factor $(n \times m)^{-1}$. Log color scale.
        (Right:) Community differences.
        Relative difference, factor between city PDFs, intents $\intents{} \times \topics{}$ topics.
        Values above 1 (blue) represent more items in Jeddah than Riyadh, likewise below 1 (pink) less. 
		\vspace{-1em}
    }
    \label{fig:confusion}
\end{figure*}

\subsection{City-Level Perspective on Jodel Content}
\label{sec:hyperlocality}

By design, Jodel users can only communicate with other Jodel users in their direct proximity---no country-wide communication is possible.
Thus, all posts carry a bias towards their local community.
We next study whether and to which extent this bias is measurable across two major cities in SA: Riyadh (the capital) vs. Jeddah about 1000\unit{km} away and believed to be more liberal.
Both largest SA cities account for 38\% (28\%, 10\%) total Jodel platform interactions in our meta data set.
As we have sampled our annotation posts accordingly, we can draw a clear picture of driving intents and discussed topics for both communities---and shifts in local topical preferences.

\afblock{Jeddah $\times$ Riyadh.}
We overall collect $\Sigma_J 1541$ ($\Sigma_R1100$) annotations for Jeddah (Riyadh) across $\Sigma 1768$ posts with substantial annotator agreement (Jeddah: $\alpha_M^\intents{}$=0.71, $\alpha_M^\topics{}$=0.62, Riyadh: $\alpha_M^\intents{}$=0.75, $\alpha_M^\topics{}$=0.63).

To ensure that we do not only observe random noise, we measure the pairwise R2 scores between per city (Jeddah, Riyadh, random) Intents $\intents{} \times \topics{}$ Topic-pdf;
resulting scores $R^2$=\{Jeddah-Riyadh: 0.75, Jeddah-random: 0.44, Riyadh-random: 0.66\} increase confidence in varying biases in discussed contents across the country.

As we are particularly interested in distribution changes, we present the relative changes between both cities as a heatmap across Intents and Topics in Figure~\ref{fig:diffs_city}. Note the log color scale.
Comparing a Jeddah Probability Distribution Functions (PDF) baseline to Riyadh results in figures above one (blue) indicating more annotations for Jeddah, and vice versa.

To draw a better picture, we added column- and row-wise averages across Intents and Topics (avg).
We find very similar intent measures in $\intents{}$-\emph{SocVentComp} $^\blacktriangleright$(factor 0.99; $\Sigma$15\% of total $\intents$-annotations). %
Non-differing topics may be considered in $\topics{}$-\emph{AnimalsNature} $^\triangleright$(0.98; $\Sigma$4\%), and \emph{Other} $^\triangleright$(1.02; $\Sigma$14\%) in 18\% overall $\topics$-annotations. %

\afblock{Content which is more popular in Jeddah (blue).}
In Jeddah, we find more $\intents{}$-\emph{Self} Expression $^\triangleright$(1.20; $\Sigma$18\%), \emph{GenEntert}ainment $^\triangleright$(1.20; $\Sigma$10\%) and individuals \emph{Seek}ing Information \& Interactions $^\triangleright$(1.23; $\Sigma$49\%) in 77\% $\intents{}$-annotations. %
Increased topic figures are rather personal and casual around $\topics{}$-\emph{SocialMedia} $^\triangleright$(1.85; $\Sigma$9\%), \emph{Self} (1.65; $\Sigma$13\%), \emph{EnterCultule} $^\triangleright$(1.19; $\Sigma$11\%), and \emph{PeopleRelation} $^\triangleright$(1.16; $\Sigma$29\%) within 63\% of total $\topics$-annotations. %

\afblock{Content which is more popular in Riyadh (pink).}
In Riyadh, we observe an overall shift towards $\intents{}$-\emph{EntObserv} $^\triangleright$(0.72; $\Sigma$3\%) and \emph{DistRelComp} $^\triangleright$(0.78; $\Sigma$14\%), both accounting for 17\% total $\intents$-anno\-ta\-ti\-ons. %
There appears to be a less heavy tailed broader spectrum of content: 
More popular topics are $\topics{}$-\emph{ProdService} $^\triangleright$(0.55; $\Sigma$6\%), \emph{IllegalViolence} $^\triangleright$(0.59; $\Sigma$2\%), \emph{BeliefsPol} $^\triangleright$(0.65; $\Sigma$11\%), \emph{FitnessHealth} $^\triangleright$(0.70; $\Sigma$4\%), \emph{EduWork} $^\triangleright$(0.78; $\Sigma$6\%), and \emph{FashionBeatuy} $^\triangleright$(0.78; $\Sigma$7\%). %
They account for another 36\% of overall $\topics$-annotations.

\takeaway{%
    By comparing the two largest---spatially distinct---com\-mu\-ni\-ties Jeddah and Riyadh, we show local biases in content occurrences.
    Riyadh experiences a broader less skewed spectrum of topics, whereas the Jeddah community focuses more on $\intents{}$-(Self Expression, Entertainment) along $\topics{}$-(Social Media, People \& Relationships).
}

\subsection{Reactions upon Content by Jodel Users}
So far, we studied what Jodel users in the KSA are talking about.
In the next step, we extend this topic with a perspective on the community reactions upon content.
That is, we raise the subsequent question whether certain intents or topics experience the same appreciation in terms of votes and replies.
Thus, we gathered the total counts of up- \& downvotes, both accumulated called Karma, and posts \& repliers per annotated thread.
From these figures, we derive two scores: 
\emph{i) Conversationness}~\cite{reelfs2022pam} as an indicator for thread participation homogeneity.
It represents the ratio of \#repliers to \#replies, \ie{} higher values equal more people with fewer responses.
And \emph{ii) Vote-consensus} as an indicator for community voting confidence.
We define it as the ratio of (\#uvotes, \#downvotes) to the total votes, while mirroring downvote-dominated values, \ie{} values approaching (negative) one depict better consistency in upvoting (downvoting) behavior; neutral at zero in between.

\begin{figure}
    \centering
    \begin{subfigure}[t]{.49\linewidth}
        \includegraphics[width=\linewidth]{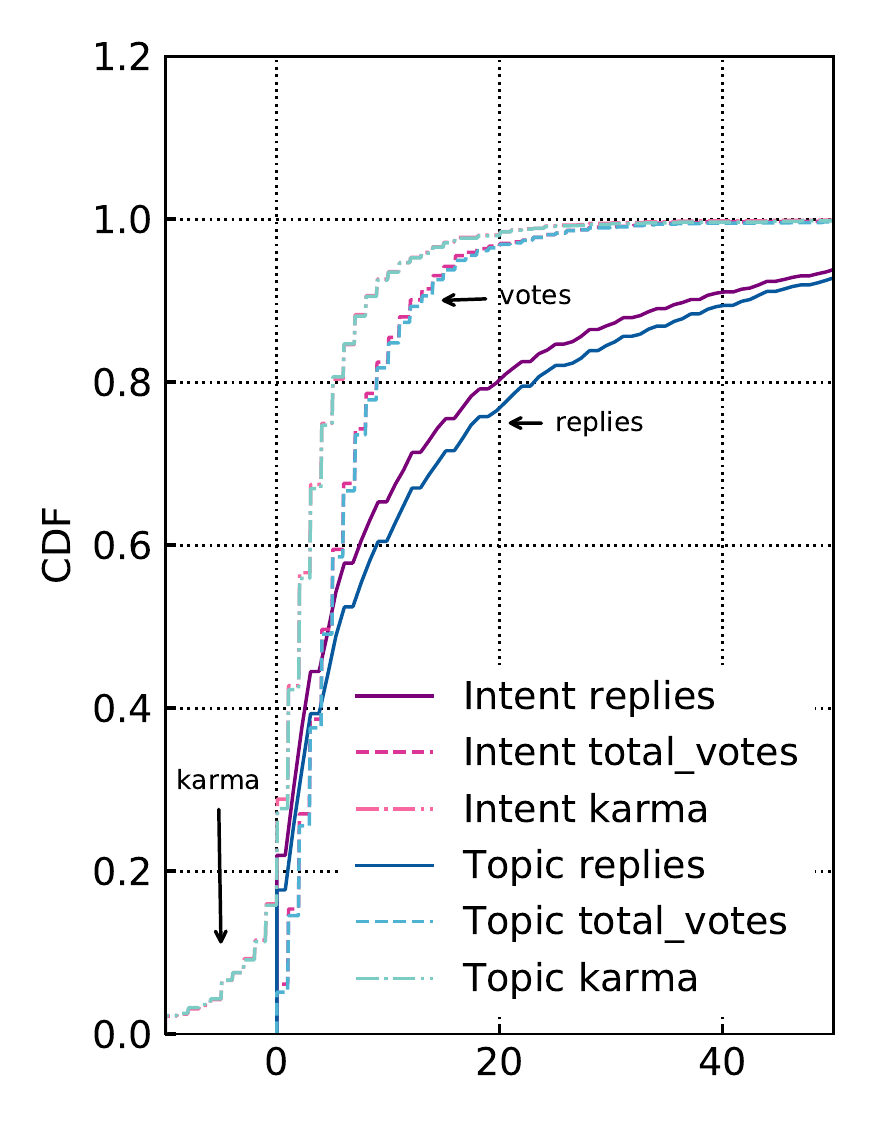}
        \caption{Replies, Votes \& Karma}
        \label{fig:metrics_cdf_replies_votes}
    \end{subfigure}
    \begin{subfigure}[t]{.49\linewidth}
        \includegraphics[width=\linewidth]{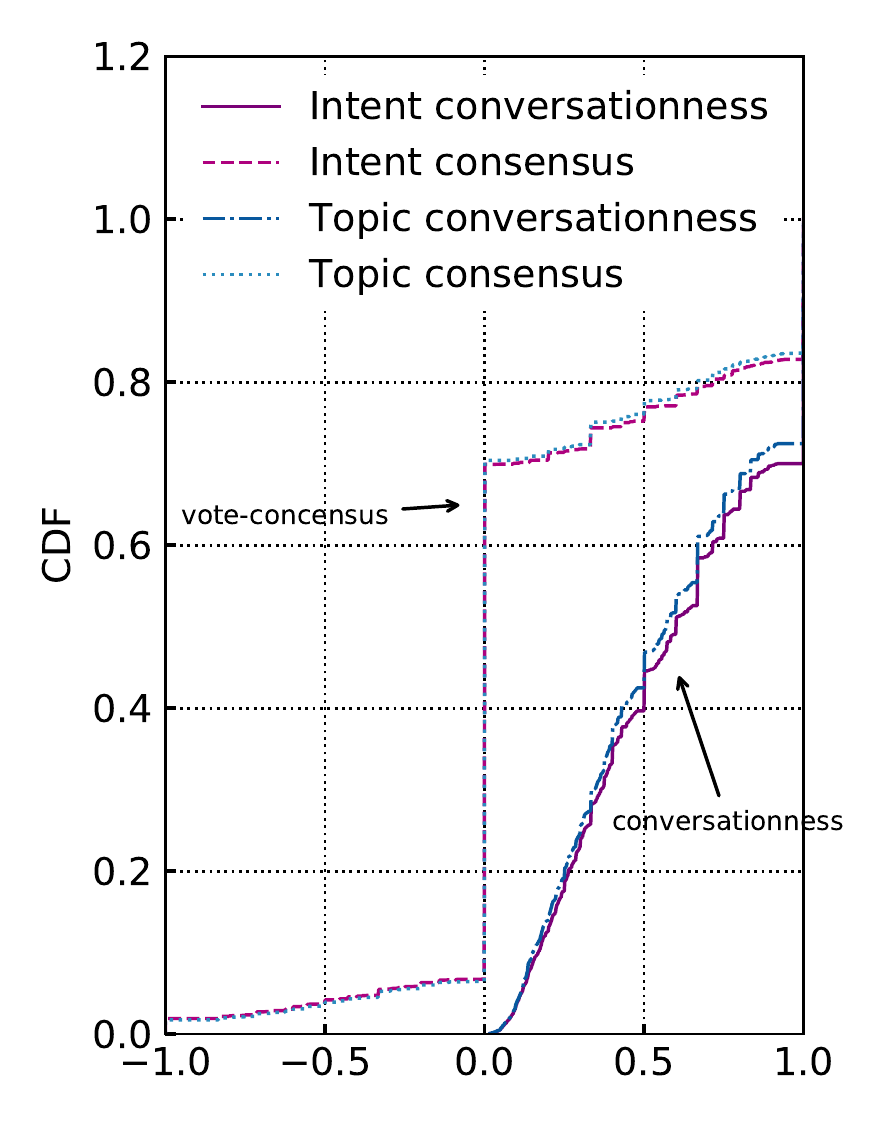}
        \caption{consensus \& Con\-ver\-sa\-t.}
        \label{fig:metrics_cdf_consensus_conversationness_new}
    \end{subfigure}
    \vspace{-1em}
    \caption{
        Received Platform Reactions.
        Vote-Related: (left)~Votes, Karma, (right)~Vote-consensus.
        Reply-Related: (left)~Replies, (right)~Conversationness.
		\vspace{-2em}
    }
    \label{fig:metrics}
\end{figure}

Buried deep within per Intent and per topic distributions, we observe similarities across most metrics; yet they differ in cut-offs or variance.
To picture an aggregated, but overall representative baseline, we present Cumulative Distribution Functions (CDFs) of Karma, replies \& votes counts in Figure~\ref{fig:metrics_cdf_replies_votes};
whereas we show CDFs of both metrics \emph{i)} \& \emph{ii)} in Figure~\ref{fig:metrics}.
To be brief, we only highlight and discuss distributional outliers.

\afblock{Replies.}
The amount of replies is generally heavy-tailed as shown in Figure~\ref{fig:metrics_cdf_replies_votes}, that might be due to app design and feed presentation; in line with other research on the structure of OSN.
However, we identify different skews within $\intents$ Intent distributions---the least replies can be expected for \emph{GenEntert} (up to 
$\uparrow$25 for 97\%), whereas others already reach up to $\uparrow$25 replies for between 80\% to 90\%.
Within $\topics$-topic distributions, the least replies can be expected for \emph{BeliefsPol} (up to $\uparrow$25 for already 90\%), appearing far less discussed than \emph{IllegalViolence}, \emph{PeopleRelation} \& \emph{FashionBeauty}  ($\uparrow$25 $\approx$70-80\%).

\afblock{Coversationness.}
In case of a low conversationness score, only few participants contribute to a long discussion.
The distributions of intents and topics are almost linear from the origin individually cutting-off ($\uparrow$x=1.0) as overall shown in Figure~\ref{fig:metrics_cdf_consensus_conversationness_new}.
Most heterogeneous discussions appear for $\intents$-\emph{Seek} \emph{Info} and \emph{SocVentComp} (cut-offs at $\uparrow$0.75 to $\uparrow$0.8), versus more homogeneous conversations in \emph{GenEntert} \& \emph{EntertObs} ($\uparrow$0.43, $\uparrow$0.58).
$\topics$-\emph{BeliefsPol} ($\uparrow$0.52) remains most homogeneously discussed at two replies per participant on average.

\afblock{Votes.}
Similar to replies, vote count distributions for Intents and topics remain heavy-tailed as shown overall series in Figure~\ref{fig:metrics_cdf_replies_votes}.

\afblock{Karma.}
Karma describes the accumulated vote score between up- and downvotes.
As can be seen in Figure~\ref{fig:metrics_cdf_replies_votes}, Jodel Karma is long-tailed to positive votes, whereas disliked posts naturally fall off around the post-remove-threshold~\cite{reelfs2022pam}.
We identify higher scores, indicating appreciation, in $\intents$-\emph{GenEntert} ($\uparrow$10 for only up to 60\%) and $\topics$-\emph{Beliefs\-Pol} ($\uparrow$10 78\%).
This evaluation also reveals that overall 17\% posts are disliked, which is slightly deceiving as most Intents and topics are significantly below this 20\% threshold.

\afblock{Vote-consensus.}
The vote consensus is bound to [-1,1], of which extremes indicate a high confidence in the community down- vs. upvotes.
As can be seen in Figure~\ref{fig:metrics_cdf_consensus_conversationness_new}, we observe an S-shape indicating that most posts experience equal up- \& downvotes (including mostly none).
However, there is an apparent skew towards positive consistency ($\approx$15\% of all posts have score 1.0).
Whereas for most $\intents$ Intents cut off the S-shape between $\downarrow$0.02 to $\downarrow$0.1 on the lower end, they reach between $\uparrow$0.61 to $\uparrow$0.79 at the upper end; $\intents$-(\emph{Other}, \emph{Info}) being outliers at (0.82, 0.88).
This general observation likewise holds true for $\topics$ topics.
Nonetheless, we find most controversially down-voted posts within $\intents$-\emph{Other} ($\downarrow$0.10, $\uparrow$0.75), whereas $\topics$-\emph{BeliefsPol} ($\downarrow$0.03, $\uparrow$0.61) are most controversial within upvotes.

\takeaway{%
    From overall distributions across the board, conversationness is rather linear with an upper cut-off, the vote-consensus is s-shaped, whereas others are commonly heavy-tailed.
    Nonetheless, we find systematic differences in community reactions upon certain $\intents{} \times \topics{}$ combinations; confirmed by only few two-sided Kol\-mo\-go\-rov-Smir\-nov tests between CDFs confidently confirming similarity over multiple metrics.
    We want to highlight various opposing outliers to $\triangle$:=($\intents$-\emph{Gen\-Entert} $\times$ $\topics$-\emph{Beliefs\-Pol})
    As for con\-ver\-sa\-ti\-on\-ness, we find most homogeneous discussions especially within $\triangle{} \cup \intents{}$-\emph{Entert\-Obs}.
    The community often appreciates $\triangle$-content with high confidence.
    In conclusion, $\triangle$ stands out: While being less discussed, discussions are more homogeneous.
    On the contrary, not spotting consistent outliers across all metrics,
    $\topics$-(\emph{Illegal\-Violence}, \emph{Social\-Media}) are discussed in longer threads with fewer participants,
    $\topics$-(\emph{Illegal\-Violence}, \emph{Prod\-Service}, \emph{Fashion\-Beauty}) can expect most replies,
    $\intents$-(\emph{Seek}, \emph{Dist\-Rel\-Comp}) $\times$ $\topics$-(\emph{Prod\-Service}, \emph{Self}, \emph{Fitness\-Health}) receive the fewest votes.
    While the latter topics are most controversially dis-/liked, the same holds true for $\intents$-\emph{Info}.
    Lastly, the communities enjoy replying to $\intents$-(\emph{Self}, \emph{Other}) $\times$ $\topics$-(\emph{Illegal\-Violence}, \emph{Fashion\-Beauty}, \emph{Fitness\-Health}).
}

\subsection{How to Not Scale Out}
Given our results, we also attempted to leverage supervised learning via SOTA pre-trained attention-based transformer masked language models, \ie{} AraBERT~\cite{antoun-etal-2020-arabert}, and used data augmentation for increasing our sample size to create automatic classifications at scale.
However, simple cross-entropy loss classification fine-tuning using an extensive hyperparameter search resulted in only 42\% accuracy for topics on our imbalanced data set.
Yet, being well \emph{above chance}, we consider achieved results as insufficient for deeper reliable insights.
Most downstream tasks base upon large amounts of data, whereas our expert classifications tend to be on the few-shot learning side.
Currently being a hot research topic, it has been shown to generally perform rather poor compared to large-scaled counterparts~\cite{few_shot_learning1__luo2021don,few_shot_learning2__wang2021grad2task}.
Thus, we believe that more data may improve results---as has our dataset room for quality improvements as well.
While computer-aided classification at large scale would be a desirable outcome, we argue that random sampling location and time creates a suitable representation of Jodel contents at substantial coder agreement to grasp users' communication intents and discussed topics.

\section{Hashtags}
\label{sec:hashtags}
After a deep dive into quantitative insights and empirically peeking into community reactions upon content, we identified driving factors intents \& discussed topics; yet we realized that it was missing crucial qualitative aspects of our annotators' experience.

\begin{table}
    \small
    \begin{tabular}{lrrr}
        \toprule
        Type                            & \#Hashtags    & \#Jodels pdf & DatingFlag \\ \midrule
        Personal Information Sharing    & 501           & 6.36\%       & 25\%       \\
        Confessions                     & 76            & 1.17\%       & -          \\
        18+                             & 70            & 1.11\%       & 74\%       \\
        Matchmaking                     & 43            & 1.02\%       & 100\%      \\
        Debates \& Opinion              & 92            & 0.86\%       & 52\%       \\
        Upvote Campaigns                & 14            & 0.55\%       & -          \\
        Other                           & 219           & 2.85\%       & 3\%        \\
        \bottomrule
    \end{tabular}
    \vspace{-1em}
    \caption{
        Classification of the top 1015 hashtags.
		\vspace{-1em}
    }
    \label{tab:hashtags}
\end{table}

\subsection{Qualitative Hashtag Classification}
That is, we now add another vector of understanding: We provide deeper insights by leveraging that hashtags inherently carry categorical information~\cite{hashtags_on_the_analysis_of__ferragina2015analyzing}.
Driven by observations and to better understand the impact of anonymity to the platform, we created a domain-specific annotation schema to capture sensitive contents and qualitatively coded the 1\unit{k} topmost hashtags accordingly.
As discussed earlier, elegantly crowdsourcing a well-suited anonymity-sensitivity score relying on many coders~\cite{wang2014whispers} is not possible in our case.
From a random 1.12\unit{M} thread subsample, we extracted all hashtags. %
Selecting the top used 1015 hashtags, we employed a single coder (age 20-25, male, Egypt) to first qualitatively screen corresponding complete conversation threads to acquire personal impressions of typical associated contents and situations.
In a next step, we created a domain-specific hashtag-annotation schema as shown in Table~\ref{tab:hashtags}, which we finally used to annotate our selected hashtags.
While we show the absolute counts of \#hashtags within each class, we also provide corresponding occurrences within the subsample (\#Jodels PDF).
According to the coder's feedback, he recognized a central recurring motif of vivid matchmaking, \emph{dating and flirting}; thus, we added a \emph{DatingFlag} correlating with this theme.

Our most prominent two hashtag categories confirm our previous findings that a user's intent often is driven by \emph{Self} or \emph{Seeking Info \& Interaction}, which is also in line with the topics \emph{Self} and \emph{People \& Relations}, or combinations of both topics with \emph{Distress Release} and \emph{Social Vententing}.
Albeit being very broad, most defined categories may be sensitive to anonymity giving \emph{Debates \& Opinions} the 18+ mark; except \emph{Upvote Campaigns}\footnote{Invitation for gathering \emph{Karma}, a lightweight in-app gamification.} and \emph{Other}.
We largely observe posts under such hashtags sharing personal experiences or confessions; other \emph{DatingFlag}ged topics often discuss love, sex, marriage, playful matchmaking, or games.

\takeaway{%
    Based on this qualitative insight, we conclude that a considerable amount of 80\% topmost hashtags relate to personal information or opinions that might not be posted in a real name environment, which is in line with previously shown main driving intents and topics---\eg{} Seeking Information \& Interaction, dating \& flirting, sharing stories and questions about People \& Relationships, while also using Jodel as a personal and social vent.
}

\subsection{Selected Taboo Picks}
Within our qualitative Hashtag study, we came across various topics, that may be considered as taboo.
That is, we find evidence for: 
\begin{itemize}
    \item Self-relief \& confessions about sexual harassment, encouraging others to share their experience (p$\approx$0.54\textperthousand),
    \item Questioning forced wear of the Niqab (p$\approx$0.54\textperthousand),
    \item Sparkling discussions about women driving (p$\approx$0.45\textperthousand),
    \item Controversial discussions and questions about homo\-sex\-uality and corresponding what-if scenarios (p$\approx$0.36\textperthousand),
    \item Words of racism against foreigners (p$\approx$0.22\textperthousand).
\end{itemize}
We find strong evidence of concrete discussed topics on Jodel within the KSA that probably would not have happened on any non-anonymous platform due to possible neglection, social pressure, and others out of manifold reasons.

\afblock{The Story of Dating \& Happy Marriages.}
In light of the main driving intent being \emph{Seeking Information and Interaction}, with Topics \wrt{} \emph{People \& Relationships} and \emph{Self Expressions}, we investigate the before mentioned recurring motif: \emph{dating \& flirting}.
Within our analyzed top 1\unit{k} hashtags, 275 were annotated with the \emph{DatingFlag} accouting for 45\unit{k} Jodels (p$\approx$4\%);
digging deeper reveals a complete storyline along getting to know each other via games, dating, questions around kissing, marriage conditions, and intercourse.

\takeaway{%
    Our observations are no exception to shown results on Whisper~\cite{wang2014whispers}.
    Also Jodel as an anonymous platform promotes sensitive content and provides a sphere where people are free in expression and more likely engage controversial discussions \& opinions---one main reason using the application as concluded from interviews~\cite{strangers_on_your_phone}.
}

Furthermore, qualitative annotator feedback concludes that Jodel also allows for any question, giving advice---or provides ventilation for personal or social distress; Yet being a source of (local) contacts, potential matches, information, good stories, and jokes.

\section{Related Work}
While a large body of work aimed at understanding Online Social Networks, two key design features of new types of networks have received little attention so far: anonymity and hyperlocality. %

Anonymous platforms are known for their ephemerally and toxicity, \eg{} 4chan~\cite{papasavva2020raiders}.
We contrast this perspective by showing that this is not generally the case; while Jodel is an anonymous platform, the posted content is largely non-toxic.
Location-basedness has been analyzed on \eg{} flickr~\cite{cha2009measurement}, or Twitter~\cite{yin2011geographical}.
The latter was used to model information diffusion~\cite{kamath2013spatio}, also conducted on Jodel itself~\cite{reelfs2019hashtag}.
Yet, Twitter and Flickr enable global communication---not being possible on Jodel---embracing local communication.

Jodel's niche of combined anonymity and hyperlocality was analyzed on \eg{} Whisper~\cite{wang2014whispers,correa2015many}, and very similar app YikYak of which we highlight a few examples.
A wide range in methodology can be found in empirical~\cite{saveski2016tracking}, but also qualitative studies~\cite{lee2017people}.
To quantify discussed topics, works leveraged (self)-supervised models~\cite{black2016anonymous} finding platform content to be rather ephemeral in an impersonal environment, which is confirmed via survey study in~\cite{vatelausyikyak,lee2017people}.
Nonetheless, another work revealed ``Personal Admission'', ``Observation'' or ``Information/Advice'' resembling popular content~\cite{heston2016invisiblecities}.
Others conducted crowdsourcing at large scale~\cite{Wu2017TPU} finding ``Dating \& Sex'' or ``Local Life, Weather \& Announcements'' being topmost discussed topics.

We conclude that research on anonymous hyperlocal platforms has matured over past years, %
yet existing studies ironically neglect their key feature---they focus on the Western/US region only.
While the Jodel app's usage has surged in the Middle East (KSA), establishing a loyal user base~\cite{reelfs21lifetime}, it is characterized by a hugely different user behavior compared to a Western counterpart~\cite{reelfs2022pam}.
Within this context, we provide answers as to what drives individual user behavior and what are discussed topics; enabled by our methodological approach to a generic crowdsourcing annotation schema.

\section{Conclusions}
We created a schema and present our methodology for assessing \emph{why} and \emph{what} humans talk about in the anonymous and hyperlocal Jodel messaging app in the Kingdom of Saudi Arabia.

Unlike common beliefs and in line with research on other anonymous location-based platforms, anonymity does not necessarily lead to toxic content at large (\eg{} hate speech).
Popular topics in Jodel focus on information seeking, entertainment, people \& relationships.
Arguably, some mentioned topics can benefit from anonymity in a society that establishes certain taboos, \eg{} casual discussions about the other sex or flirting.
An anonymous platform can support such topics and enable an atmosphere in which users are free in their expressions as also shown in~\cite{correa2015many}.
What they discuss can differ between cities, as shown by comparing Riyadh and Jeddah, with Riyadh having a broader spectrum of topics available.
By evaluating votes (content appreciation) and replies (reactions), we show that the communities react differently to different topics; \eg{} entertaining posts are much appreciated through votes, receiving the least replies, while beliefs \& politics receive similarly few replies but are controversially voted.

Our study shows a lower-bound on the prevalent topics in an anonymous and hyperlocal messaging app.
Our classification scheme enables future work to assess topical preferences more broadly.

\emph{
``As for love, it still might always struggle to come out into the light of day in Saudi Arabia.
You can sense that in the sighs of bored men sitting alone in cafés, in the shining eyes of veiled women walking down the streets [...], and in the heartbroken songs and poems, too numerous to count, written by the victims of love unsanctioned by family, by tradition, by the city: Riyadh''
}
~\cite[pp. 313-314]{girlsOfRiyadh}.

\begin{acks} 
  We are deeply thankful to all our students for their exceptional effort enabling this study: 
  Ahmed Soliman,
  Haitham Almasri,
  Hana Al Raisi,
  Dalia \& Mahmoud Moussa,
  Marcel Kröker,
  and Shams Dulaimi.
\end{acks}

\pagebreak

\balance
\bibliographystyle{ACM-Reference-Format}
\bibliography{references}

\end{document}